\documentclass[lettersize,journal]{IEEEtran}
\usepackage{amsmath,amsfonts}
\usepackage{algorithmic}
\usepackage{algorithm}
\usepackage{array}
\usepackage[caption=false,font=normalsize,labelfont=sf,textfont=sf]{subfig}
\usepackage{url}
\usepackage{verbatim}
\usepackage{graphicx}
\usepackage{cite}
\usepackage{amsmath,amssymb,amsfonts}
\usepackage{graphicx}
\usepackage{textcomp}
\usepackage{multirow}
\usepackage{caption}
\usepackage{threeparttable}
\usepackage{stfloats}
\usepackage{epstopdf}
\begin{document}
\title{Device Fault Prediction Model based on LSTM and Random Forest}
\author{Jing Xu and Yongbo Zhang
         \thanks{ }
}

\markboth{Journal of \LaTeX\ Class Files,~Vol.~XX, No.~X, June~2023}%
{Shell \MakeLowercase{\textit{et al.}}: A Sample Article Using IEEEtran.cls for IEEE Journals}


\maketitle

\begin{abstract}
  The quality of power grid equipment forms the material foundation for the safety of the large power grid. Ensuring the quality of equipment entering the grid is a core task in material management. Currently, the inspection of incoming materials involves the generation of sampling plans, sampling, sealing, sample delivery, and testing. Due to the lack of a comprehensive control system and effective control measures, it is not possible to trace the execution process of business operations afterward. This inability to trace hampers the investigation of testing issues and risk control, as it lacks effective data support. Additionally, a significant amount of original record information for key parameters in the testing process, which is based on sampling operation standards, has not been effectively utilized. To address these issues, we conduct researches on key monitoring technologies in the typical material inspection process based on the Internet of Things (IoT) and analyze the key parameters in inspection results. For purpose of complete the above tasks, this paper investigates the use of Long Short-Term Memory (LSTM) algorithms for quality prediction in material equipment based on key inspection parameters. In summary, this paper aims to provide professional and reliable quality data support for various business processes within the company, including material procurement, engineering construction, and equipment operation.
\end{abstract}

\begin{IEEEkeywords}
Equipment Failure Prediction, Fault Diagnosis, Anomaly Detection
\end{IEEEkeywords}

\section{Introduction}
\IEEEPARstart{T}{his} paper primarily focuses on equipment quality inspection of surge arresters. As an essential overvoltage protection devices in power systems, surge arresters are designed to suppress overvoltages (including operational overvoltages and lightning-induced overvoltages), release high currents, and prevent protected equipment from being damaged by lightning-induced overvoltages or operational overvoltages \cite{durbak2001surge, ranjbar2022survey, castro2022optimal}. Surge arresters not only affect the ability of power systems to operate safely and stably but also impact their economic efficiency, especially in the construction of extra-high voltage and ultra-high voltage systems. They serve as a robust safeguard for the secure operation of power systems \cite{pinceti1999simplified}.

Currently, metal oxide surge arresters (MOAs) have largely replaced traditional surge arresters (such as silicon carbide valve-type surge arresters) \cite{falcaro2016application,das2021transfer}.  While MOAs have eliminated some of the drawbacks of traditional surge arresters and offer superior impulse current withstand capabilities, they still have certain limitations. Firstly, due to the elimination of series gaps in MOAs, there is some leakage current through their valve elements. This results in cumulative thermal effects and aging of the valve elements, which can even lead to thermal breakdown and explosions \cite{tominaga1979protective}.  Secondly, after many years of operation, exposure to high-energy impulse currents can cause aging \cite{thomas1970high}.  Lastly, external factors like moisture or contamination can lead to increased leakage current or decreased insulation resistance, and increased grid frequency current can cause abnormal internal discharges \cite{borges2021review}.

During operation, surge arresters can develop defects like internal insulation moisture and aging of valve elements  \cite{kester1998multistress,das2021novel}. These defects not only reduce the ability of protected equipment to withstand overvoltages but also  pose risks to personnel and affect the safe operation of the power system. Traditional approaches aim to conduct regular inspections and maintenance, thus enabling the grid to stably operate, ensuring the safety of the protected equipment, and preventing  surge arresters from some serious grid accidents due to their own failures. Suspicion of defects in MOAs can be addressed by preventive tests to check their operational status and ensure that defects have been detected and eliminated promptly. Regular tests require the shutdown of protected equipment to prevent damages to its insulation performance. Therefore, the test conditions are usually not representative of the actual operating conditions of the surge arrester, and the test results cannot fully reflect its actual condition during operation. Furthermore, the methods of regular testing cannot prevent faults that occur between two test intervals.

To overcome these drawbacks, the industry widely adopts online monitoring of surge arresters \cite{metwally2017online,barannik2020system,abdullah2023surge}. This involves directly monitoring key parameters (such as leakage current, resistive current, resistive current fundamental frequency, and resistive current third harmonic) of MOAs. Online monitoring systems assess the operating status and the degree of defects directly through simple threshold comparisons or, more indirectly, by using more accurate and forward-looking algorithms. Compared to traditional blackout testing, online monitoring has several advantages: it does not require a power outage, does not cause a power interruption, provides real-time monitoring of the MOA's operating status, allows for timely detection of defects and failures, and reduces the routine investment of human, material, and time resources. In addition, online monitoring systems collect massive operational data of MOAs during their operation. These historical data can be deeply analyzed and mined to improve judgment accuracy, which is also significant for the development of big data in smart grids. Hence, implementing online monitoring of MOAs, conducting quasi-real-time or offline status assessments, issuing early defect warnings, and considering the long-term operation and development of the power grid are all valuable approaches.

Online monitoring theoretically reflects the fault status of surge arresters to some extent \cite{hinrichsen1999online}. The complexity of surge arresters' working environment, along with environmental factors like temperature, humidity, and surface contamination, significantly reduces the accuracy of measured data. As a result, relying solely on measured values becomes challenging for making final judgments. The accurate judgment of the internal operating status of surge arresters requires combining other relevant variables.

This paper primarily focuses on the construction of a surge arrester fault prediction model based on LSTM and random forest algorithms, in combination with online monitoring data and the environmental information. The evaluation results of the model are of great significance for evaluating the state of the surge arresters and finding its defects in time, which can maximize the lightning protection function of the surge arresters. This method can enriches and improves the functions of the MOA online monitoring system. The developed fault prediction algorithm based on LSTM and random forest can also be used for online monitoring and analysis warning of other equipments in the power system, such as transformers and transmission lines. It has portable and universal significance.

\section{Related Work}
The research approaches for assessing the status and analyzing defects of surge arresters primarily include the following methods: measurement and analysis of individual features (leakage current, its resistive component, harmonic components, etc.); improvement and optimization of measurement methods for these features (noise and interference elimination, efficient information extraction from data, etc.); simple consideration and analysis of multiple features in combination; application of algorithmic models (traditional algorithms and machine learning algorithms) based on data from multiple features.

Lundquist \textit{et al.} \cite{lundquist1990new} introduced a novel approach, the Capacitance Bridge method, to measure the leakage current resistance of MOAs. The key point of the Capacitance Bridge method is to use capacitive current as a compensating reference signal. Experimental results demonstrate the significant advantages of this method, as it greatly reduces measurement interference caused by harmonics, simplifies leakage current data measurement, and allows for waveform observation and comparison on an oscilloscope. Cai \textit{et al.} \cite{cai2010line} developed a mathematical morphological technique to eliminate signal noise and subsequently deduce the resistive fundamental frequency leakage current through signal correlation analysis. This method effectively enhances the accuracy of MOA leakage current detection, ensuring the effectiveness and accuracy of subsequent MOA defect diagnosis.

Hu \textit{et al.} \cite{hu2019operation} investigated an online status monitoring system for surge arresters based on multi-layer vector machines. It conducted tests and comparisons between surge arresters with various defects (such as aging due to solar radiation, aging of resistive valve elements, and contamination). Similarly, Lee \textit{et al.} \cite{lee2022universal} proposed a fuzzy environment-based MOA fault identification method. It locally fuzzified the surrounding environment affecting leakage current and used an adaptive particle swarm algorithm to obtain optimal values for the SVM penalty factor and kernel function parameters to achieve the best performance.

Based on the above research, this study develops a surge arrester fault warning technology based on LSTM and random forest algorithms. It initially utilizes a time series-based LSTM model to perform pre-analysis of faults and their development. Subsequently, it employs the random forest method from machine learning to analyze the types of faults.


\begin{figure*}[ht]
{\centering
\includegraphics[scale=0.65]{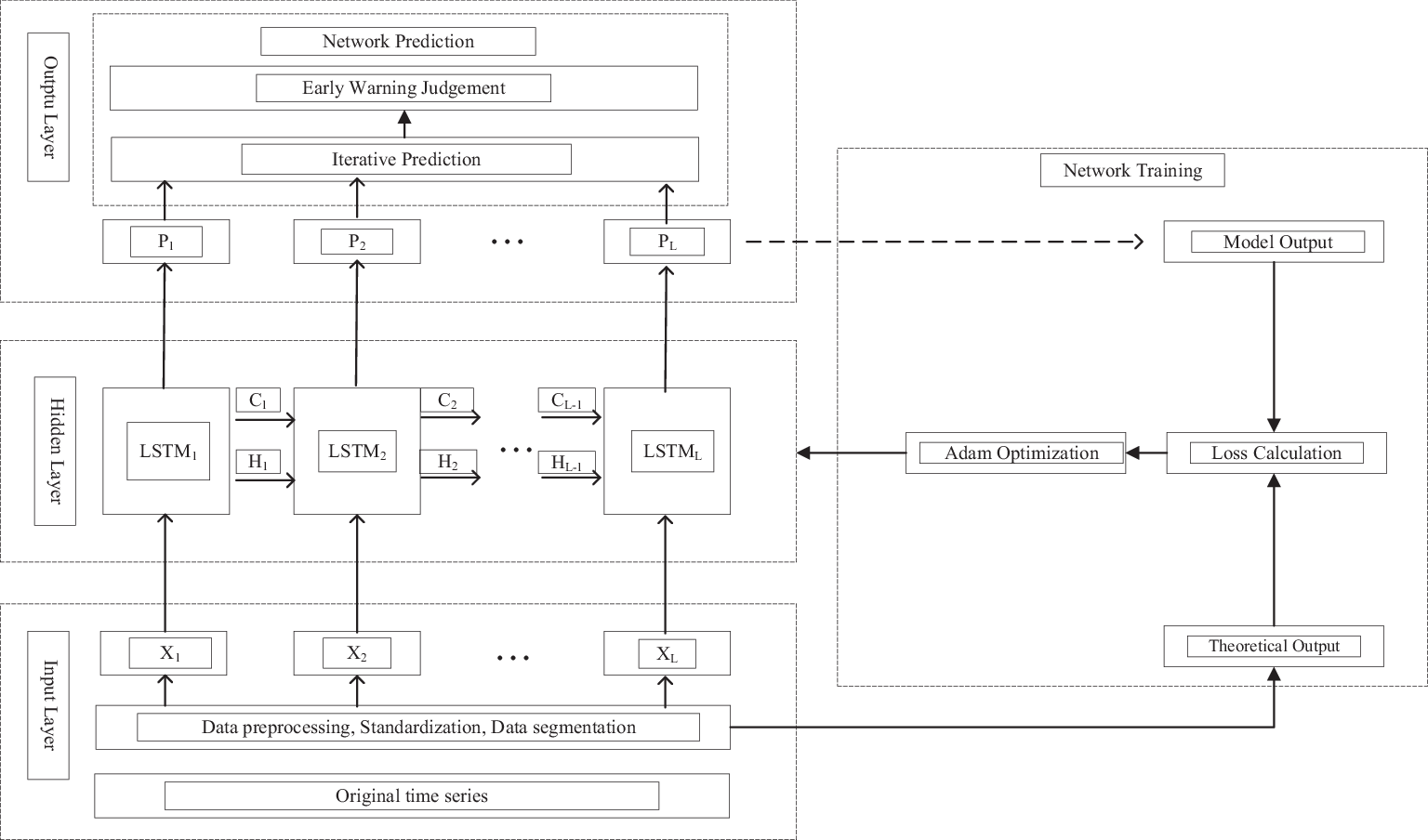}
\caption{Time series prediction framework based on LSTM}
\label{fig1}}
\end{figure*}

\section{Methods}
\subsection{LSTM-based device future state warning algorithm}
A time series of a random variable refers to a sequential curve representing the variation of the variable over time. Delving into the essence of a variable's time series can unveil the fundamental patterns of how it changes over time. Following these patterns enable the prediction of the variable's evolution over a future period. For power grid equipment, aging or latent defects often develop slowly and exhibit strong concealment. Relying solely on threshold-based logic for monitoring may struggle to identify situations where key feature data, although not exceeding limits, show a steady increase. Therefore, in this study, we employ trend prediction algorithms based on time series to issue warnings in such cases. The algorithm forecasts the value of key feature variables of power grid equipment for the next moment and compares it with a predefined threshold. When the predicted data surpasses the threshold, the equipment's state is flagged as potentially abnormal.

By collecting real-time monitoring data of power grid equipments and developing an intelligent predictive warning algorithm based on time series and trend analysis, this research aims to provide early warnings for equipment defects. Considering the limited sample points in the time series of a single-variable monitoring feature, as well as the principle of designing a simplified recurrent neural network, the overall framework of constructing the LSTM prediction model is depicted in Fig.1. It consists of five functional modules: the input layer, hidden layers, the output layer, network training, and network prediction. The input layer preprocesses the time series of the original monitoring feature variable to meet the network's input requirements \cite{yu2019review, smagulova2019survey}. The hidden layers employ a two-layer recurrent neural network structured with LSTM cells. The output layer generates predictive results and assesses them for warning decisions. Network training utilizes the Adam optimization algorithm, while network prediction employs an iterative approach for point-by-point forecasting.

\begin{figure*}[ht]
{\centering
\includegraphics[scale=0.65]{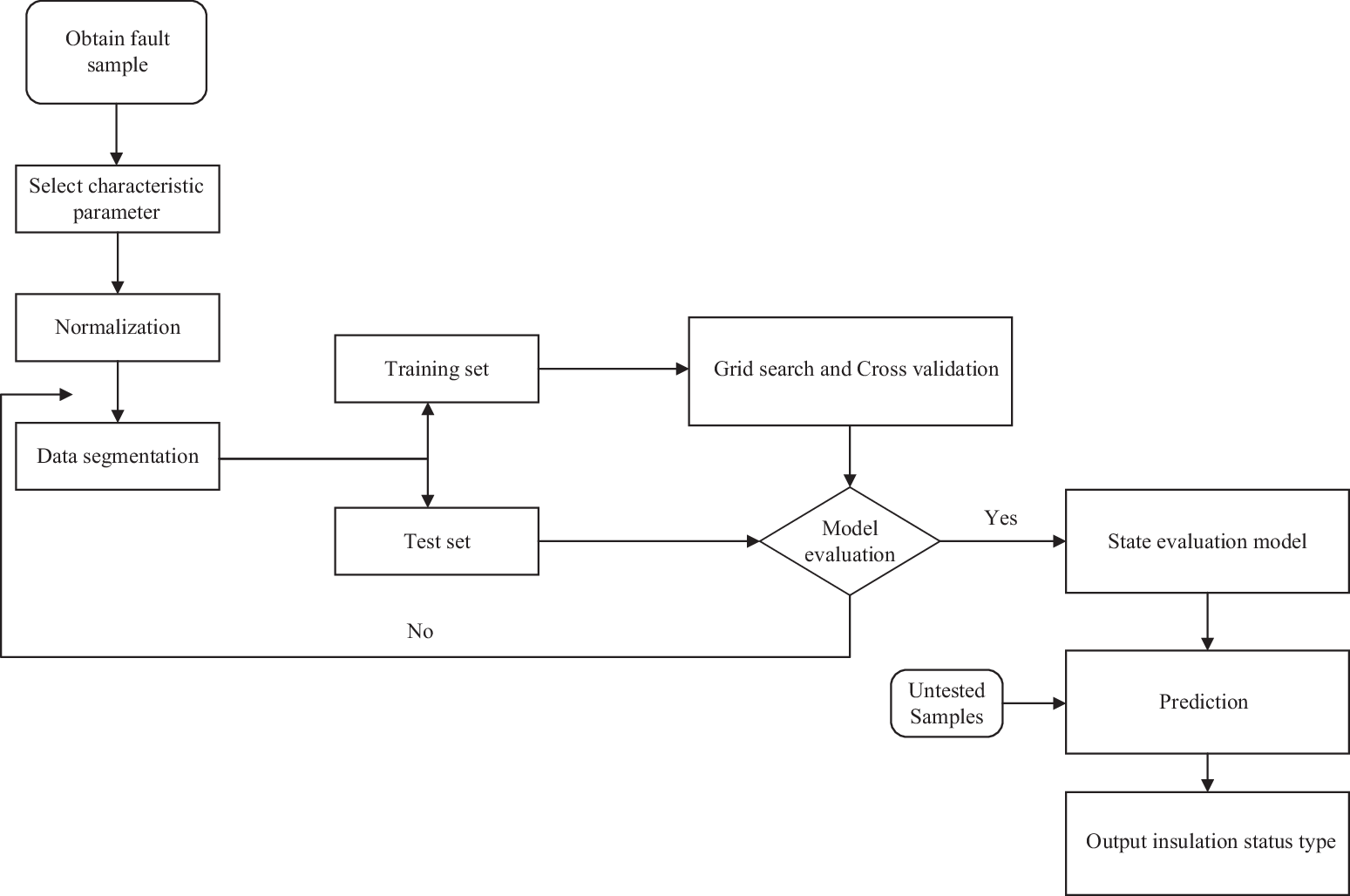}
\caption{Insulation state assessment based on random forest}
\label{fig1}}
\end{figure*}

\subsection{Equipment fault type recognition model based on random forest}
This section is based on the feature data collected from the monitored power grid equipment. It involves analyzing the equipment's state or defects by examining the values of various features. The underlying principle is to construct a Random Forest (RF) model based on known feature values and fault types \cite{biau2016random}. This model is used to calculate the probability of each fault type for power grid equipment. The fault type corresponding to the highest probability is selected as the predicted fault type.

In the classification of power grid equipment states, distinguishing each healthy state can be viewed as a binary classification process. The Area Under the Curve (AUC) is a crucial metric for assessing the performance of binary classifiers. Consequently, in this paper, the AUC metric is integrated with the RF algorithm. This integration involves computing the AUC values for each decision tree to assign different voting weights to each classifier. The improved RF algorithm's workflow is illustrated in Fig.2.

\section{Experiments}
\subsection{Method Comparison}
To validate the high predictive accuracy of the LSTM prediction model established in this paper, the ARIMA \cite{piccolo1990distance} method was employed to construct prediction models for the monitored variables, leakage current, and resistive current of the surge arrester using the same dataset. ARIMA is a classical theory and method in time series analysis, and its model can be represented as ARIMA(p, d, q), where p, d, and q represent the number of autoregressive terms, the order of differencing, and the number of moving average terms, respectively. In practical usage, these three parameters can be determined through observing the Autocorrelation Function (ACF) and Partial Autocorrelation Function (PACF) or by calculating the Akaike Information Criterion (AIC) or Bayesian Information Criterion (BIC) values to find the minimum.

Using the constructed ARIMA and LSTM prediction models, the prediction performance of the surge arrester's leakage current and resistive current was evaluated on the same test dataset. The assessment was conducted using the two metrics, namely, the Mean Absolute Percentage Error (MAPE) and the Root Mean Square Error (RMSE). Smaller values of these two metrics indicate better prediction accuracy. The results of comparing the ARIMA and LSTM models for predicting the surge arrester's leakage current and resistive current are presented in Table \uppercase\expandafter{\romannumeral1}.

\begin{table}[htbp]\centering \small
\captionsetup{font={small}}
\renewcommand{\arraystretch}{1.5}
\setlength{\tabcolsep}{4mm}
\setlength{\belowcaptionskip}{2pt}
\caption{ Prediction accuracy on ARIMA and LSTM}
\scalebox{1}{

    \begin{threeparttable}
    \resizebox{\linewidth}{!}{
        \begin{tabular}{c|c|c|c}
          \hline
          \textbf{Variables} & \textbf{Methods} & \textbf{MAPE -- \% }& \textbf{RMSE -- \% }\\
          \hline

          Full leakage current & LSTM & 5.62 & 0.154\\

          Full leakage current  & ARIMA & 6.82 & 0.171\\

          Resistive current & LSTM & 4.16 & 0.136 \\

          Resistive current & ARIMA & 5.42 & 0.162\\
          \hline
        \end{tabular}
        }
  \end{threeparttable}
  }
\end{table}

From Table \uppercase\expandafter{\romannumeral1}, it can be observed that in the prediction datasets from groups 1 to 12, the LSTM model exhibits higher prediction accuracy and better prediction stability compared to the ARIMA model. The LSTM model has MAPE values of 5.62\% and 4.16\% and RMSE values of only 0.154 and 0.136. In contrast, the ARIMA model exhibits relatively lower accuracy with MAPE values of 6.82\% and 5.42\%, indicating larger prediction errors. This is particularly noticeable during the troughs, where a significant gap exists between the predicted and actual values, and the predictions lag behind the actual values.

Comparing the two methods, LSTM and ARIMA, it can be concluded that LSTM outperforms ARIMA in terms of precision and fitting. LSTM is better at handling the lag in the changes of leakage current and resistive current in uncertain environmental conditions, achieving higher predictive accuracy. This demonstrates the strong adaptability of LSTM in predicting surge arrester monitoring variables.

\subsection{Equipment fault type recognition model based on random forest}

\begin{table}[htbp]\centering \tiny
\renewcommand{\arraystretch}{1.5}
\captionsetup{font={small}}
\setlength{\tabcolsep}{4mm}
\setlength{\belowcaptionskip}{2pt}
\caption{ Prediction results of random forest model}
\scalebox{1}{
    \begin{threeparttable}
    \resizebox{\linewidth}{!}{
        \begin{tabular}{c|c|c}
          \hline
          \textbf{Variables} & \textbf{Percision -- \% } & \textbf{Recall -- \% }\\
          \hline
          Normal & 93.4 & 92.5 \\
          Aging & 84.3 & 85.5 \\
          Damp & 83.2 & 78.5 \\
          Surface & 80.1 & 79.2 \\
          Other causes & 82.1 & 80.6 \\
          \hline
        \end{tabular}
        }
  \end{threeparttable}
  }
\end{table}

Finally, based on the test dataset, an evaluation of the trained model was conducted using metrics such as overall accuracies, recall, and precision. From the prediction results for various insulation states as shown in Table \uppercase\expandafter{\romannumeral2}, we can see that the normal state has the highest recall and precision rates, while the recall and precision rates for other fault types exceed 78\%. The model demonstrates a strong predictive capability for various fault categories. Furthermore, the model achieves an average prediction accuracy of 92.6\%. To further validate the superiority of the random forest model, it was compared with models constructed using decision trees and support vector machines. This comparison was repeated 10 times, with 80\% of the original data randomly selected as training samples each time. Random forest, decision tree, and support vector machine models were built using the training samples, while the remaining 20\% of the data served as the test samples. The prediction accuracy of different models was calculated. The average overall accuracy of the random forest model, decision tree model, and support vector machine model are 92.6\%, 85.23\%, and 84.75\%, respectively. Consequently, the random forest model exhibits superior generalization and higher prediction accuracy. Experimental results show that the parameter optimization based on a multi-grid search helps improve the model's prediction performance. Compared to the typical time series forecasting model ARIMA, the LSTM model has overall better fitting and prediction performance, with predicted trends and fluctuations closely aligning with the actual values. When the surge arrester warning method based on LSTM issues a fault prompt or alarm, a subsequent step involves determining whether the alarm is a false positive and identifying the specific fault type. This is accomplished by inputting the predicted data from LSTM into a pre-trained random forest model for fault type prediction. This model combines multiple pieces of information, such as surge arrester online monitoring and environmental factors, considering the most typical feature variables representing the operating state. It has high credibility, tolerance, and adaptability. It also has characteristics of fast recognition, stable learning algorithms, and ease of training compared to other state assessment methods. Actual test data shows that the random forest model established in this study has a higher average prediction accuracy than that of decision tree and support vector machine models, indicating its strong reliability.

\section{Conclusion}
In theory, traditional methods for diagnosing surge arresters can reflect the arrester’s fault status to some extent. However, due to the influence of external factors such as temperature, humidity, and contamination, the measured quantities are difficult to use as the final basis for judgment. In this paper, a surge arrester insulation state warning model based on LSTM and random forest algorithms is proposed. The model consists of two main steps: Firstly, a deep neural network algorithm, LSTM, suitable for time series data is applied. By training historical data, it predicts the future state of key features of the surge arrester. If the prediction result (or growth trend) exceeds a set threshold, it issues a fault prompt or alarm. Then, the predicted data from LSTM will be input into a pre-trained random forest model for fault type prediction, determining the specific fault type. Since the degradation or latent defects of surge arresters develop slowly, it is difficult to identify abnormal states when the monitoring data of the surge arrester does not exceed the limit but steadily increases, based solely on threshold values. The early warning method of lightning arrester based on LSTM recurrent neural network has the ability to predict the leakage current data of lightning arrester at the next time. It can longitudinally analyze the changing trend of the current and compare it with the set threshold. When the predicted data exceeds the threshold, it will issue a warning of abnormal surge discharger status. The algorithm includes training, prediction and parameter optimization of LSTM model.

\bibliographystyle{IEEEtran}
\bibliography{paper}


\vfill

\end{document}